\documentclass{PoS}
\usepackage{amsmath,amssymb}

\newcommand{\bm}[1]{\vec{#1}}
\newcommand{\e}{\mbox{e}}

\title{Spectral properties of quarks above $T_c$ 
-- thermal mass, dispersion relation, and self-energy --}

\ShortTitle{Spectral properties of quarks above $T_c$}

\author{\speaker{Masakiyo Kitazawa}\\
Department of Physics, Osaka University, Toyonaka, Osaka, 560-0043, Japan\\
E-mail: \email{kitazawa@phys.sci.osaka-u.ac.jp}}

\abstract{
Spectral properties of quarks above the critical temperature for
deconfinement are analyzed in quenched lattice QCD on lattices of
size $128^3\times16$.
We study quark spectral function in energy and momentum space,
focusing on the values of the thermal mass and the dispersion
relations of normal and plasmino modes at nonzero momentum, as well
as their spatial volume dependence. Our numerical result suggests
that the dispersion relation of the plasmino mode has a minimum at
nonzero momentum even near the critical temperature. 
The quark self-energy is also analyzed by using the analyticy of
the inverse propagator, which is found to be consistent with the 
spectral function estimated by the two-pole ansatz.
}

\FullConference{The XXVIII International Symposium on Lattice Field Theory, Lattice2010\\
		June 14-19, 2010\\
		Villasimius, Italy}

\begin{document}

\section{Introduction}

At asymptotically high temperature ($T$), properties of strongly 
interacting matter described by Quantum Chromodynamics (QCD) 
can be calculated using perturbative techniques. 
In this limit, it is known that the collective excitations of 
quarks develop a mass gap (thermal mass) and a decay rate 
proportional to $gT$ and $g^2T$, respectively, where $g$ 
denotes the gauge coupling \cite{LeBellac}.
Furthermore, the dispersion relation splits into two branches,
the normal and plasmino modes.
As the decay rate parametrically grows faster than the thermal
mass as $g$ increases, it is na\"ively expected that the quark 
quasi-particles cease to exist as $T$ is lowered.
On the other hand, quark number scaling of the elliptic flow 
observed in RHIC experiments indicates the existence of 
quasi-particles having a quark quantum number \cite{Fries:2003kq}.
To understand properties of the matter near $T_c$, especially
the quasi-particle nature of elementary excitations,
therefore, it is desirable to explore the spectral properties
of quarks within nonperturbative techniques. 


Recently, the correlation function of quarks at nonzero $T$ 
has been analyzed on the quenched lattice with size up to 
$N_\sigma^3 \times N_\tau = 64^3\times16$ in Landau gauge \cite{KK}.
In these studies, it is found that the quark correlation 
function above $T_c$ obtained on the lattice is well reproduced 
by the two-pole ansatz for the spectral function, where the 
two poles correspond to the normal and plasmino modes.
In the chiral limit these modes have identical quasi-particle 
masses, which are identified to be the thermal mass, that are 
approximately proportional to $T$. 
These calculations, however, also showed that the 
quasi-particle masses, which perturbatively arise through the 
resummation of infra-red sensitive loops 
\cite{LeBellac}, are strongly dependent on the
physical volume, $V$, used in the lattice calculations.

The purpose of the present study is to extend the analysis 
in Refs.~\cite{KK} to much larger spatial volume, 
$N_\sigma^3 \times N_\tau = 128^3\times16$ \cite{KKKS}.
This analysis enables us to investigate the spatial volume 
dependence in more detail. The large spatial volume also 
allows to directly analyze the momentum dependence of excitation 
spectra, i.e. the dispersion relations, more precisely.
To obtain better understanding on the spectral property
of quarks, in addition to the analysis of the spectral
function, we also try to examine the quark self-energy by 
numerically taking the inverse of the quark propagator.

\section{Quark spectral function and fitting ansatz}

Excitation properties of the quark field are encoded in 
the quark spectral function $\rho_{\mu\nu}(\omega,\bm{p})$, 
with $\mu$ and $\nu$ denoting Dirac indices.
The Dirac structure of $\rho_{\mu\nu}(\omega,\bm{p})$ 
at finite temperature is decomposed as
\begin{align}
\rho_{\mu\nu}( \omega, \bm{p} )
= \rho_0( \omega,p ) (\gamma^0)_{\mu\nu}
- \rho_{\rm v}( \omega,p ) (\hat{\bm{p}}\cdot\bm{\gamma})_{\mu\nu}
+ \rho_{\rm s}( \omega,p ) 1_{\mu\nu},
\label{eq:rho_0vs}
\end{align}
where $p=|\bm{p}|$ and $\hat{\bm{p}}=\bm{p}/p$.
In the present study we consider the spectral function above $T_c$
for two cases; (1) at zero momentum, and 
(2) in the chiral limit.
In these cases, the Dirac structure of $\rho_{\mu\nu}(\omega,\bm{p})$ 
are decomposed by using projection operators
\cite{KK}.
With $p=0$, $\rho_{\rm v}( \omega,p )$ vanishes in 
Eq.~(\ref{eq:rho_0vs}) and $\rho_{\mu\nu}( \omega,\bm{p}=\bm{0} )$ 
is decomposed with the projection operators 
$L_\pm = ( 1 \pm \gamma^0 )/2$ as 
\begin{align}
\rho( \omega, \bm{0} )
= \rho^{\rm M}_+( \omega ) L_+ \gamma^0 
+ \rho^{\rm M}_-( \omega ) L_- \gamma^0.
\label{eq:rho^M}
\end{align}
In the chiral limit and for $T>T_c$, the system possesses 
the chiral symmetry and $\rho_{\rm s}( \omega,p )$ vanishes.
In this case, $\rho_{\mu\nu}( \omega,\bm{p} )$ is 
decomposed with the projection operators 
$P_\pm (\bm{p})= ( 1 \pm \gamma^0\hat{\bm{p}}\cdot\bm{\gamma} )/2$
as 
\begin{align}
\rho( \omega,\bm{p} )
= \rho^{\rm P}_+( \omega,p ) P_+(\bm{p}) \gamma^0
+ \rho^{\rm P}_-( \omega,p ) P_-(\bm{p}) \gamma^0.
\label{eq:rho^P}
\end{align}

In order to extract $\rho_{\mu\nu}(\omega,\bm{p})$ from lattice 
QCD simulations we have analyzed the quark correlation function 
in Euclidean space
\begin{align}
S_{\mu\nu}( \tau,\bm{p} )
= \frac1V \int d^3x d^3y 
\e^{ i {\bf p} \cdot ( {\bf x}-{\bf y} ) }
\langle \psi_\mu( \tau,\bm{x} ) \bar\psi_\nu ( 0,\bm{y} ) 
\rangle,
\label{eq:S}
\end{align}
on the lattice in quenched approximation in Landau gauge.
Nonperturbatively-improved clover fermion is used for the analysis.
Equation~(\ref{eq:S}) is related to
the spectral function as
\begin{align}
S_{\mu\nu}( \tau,\bm{p} )
= \int_{-\infty}^\infty d\omega
\frac{ \e^{ (1/2 -\tau T) \omega/T }}{ \e^{\omega/2T} + \e^{-\omega/2T} }
\rho_{\mu\nu}( \omega,\bm{p} ).
\label{eq:Stau-rho}
\end{align}

In order to examine the quark spectral function from 
lattice correlator, we follow the approach taken in 
\cite{KK}, which makes use of a two-pole ansatz for the 
spectrum,
\begin{align}
\rho^{\rm M,P}_+(\omega)
= Z_1 \delta( \omega - E_1 ) + Z_2 \delta( \omega + E_2 ).
\label{eq:2pole}
\end{align}
Here, $Z_{1,2}$, and $E_{1,2}>0$, are fitting parameters that
will be determined from correlated fits to the lattice 
correlator: $Z_{1,2}$ and $E_{1,2}$ represent the 
residues and positions of poles, respectively. When comparing
the fit results with spectral functions obtained in 
perturbative calculations one can identify the pole at 
$\omega=E_1$ to be the normal mode, while the one at 
$\omega=-E_2$ corresponds to the plasmino mode \cite{BBS92,KKN}.
To determine the fit parameters 
with correlated fits, we use lattice data points at
$\tau_{\rm min} \le \tau/a \le N_\tau -\tau_{\rm min}$ 
with $\tau_{\rm min}=4$.
We found that the ansatz Eq.~(\ref{eq:2pole}) gives
a reasonable chi-square, $\chi^2/{\rm dof}\simeq1$, 
over wide parameter ranges \cite{KK}; see, however, 
Ref.~\cite{KKKS} which discusses problems associated with 
the analysis of the spectral function with an ansatz.

\section{Thermal mass and dispersion relations}

\begin{figure}[tbp]
\begin{center}
\includegraphics[width=.55\textwidth]{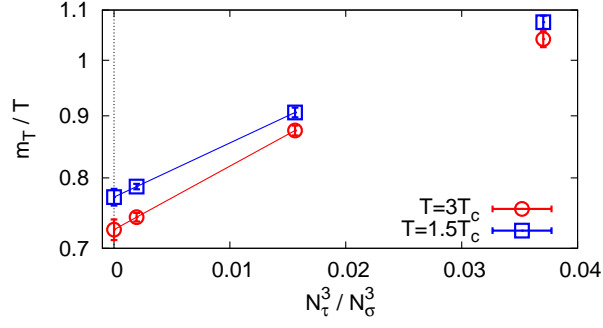}
\caption{
Extrapolation of the thermal mass of the quark $m_T$ 
obtained by the pole ansatz to the infinite volume limit for 
$T/T_c=3$, $1.5$.
}
\label{fig:extrpl}
\end{center}
\end{figure}

The quark spectral function is analyzed 
(1) for $p=0$ as a fuctions of bare quark mass, $m_0$, and
(2) for $m_0=0$ as a fuctions of $p$.
For each case, we use the projection Eqs.~(\ref{eq:rho^M})
and (\ref{eq:rho^P}), respectively.

In the analysis of $\rho^{\rm M}_\pm(\omega)$ at $p=0$ on the 
largest lattice with $N_\sigma/N_\tau=8$, we found that the 
$m_0$ dependence of fitting parameters qualitatively agrees 
with previous results obtained on lattices with 
$N_\sigma/N_\tau=4$ and $3$ in Ref.~\cite{KK}.
From the $m_0$ dependence of $\rho^{\rm M}_\pm(\omega)$ one
can define the critical hopping parameter for the chiral limit 
and the thermal mass, $m_T$ \cite{KK,KKKS}.
In Fig.~\ref{fig:extrpl}, we show the value of $m_T$ with 
$N_\sigma/N_\tau=8$, $4$ and $3$ for $T/T_c=1.5$ and $3$.
To infer the thermal mass in the infinite volume limit, 
we performed an extrapolation of $m_T$ to infinite volume 
with an ansatz $m_T(1/V) \sim m_T(0) \exp(c/V)$ using the 
results with $N_\sigma/N_\tau=8$ and $4$.
The result of this extrapolation is shown in 
Fig.~\ref{fig:extrpl}.
Within the extrapolation, the value of $m_T(1/V\to0)$ 
coincides the one obtained with $N_\sigma/N_\tau=8$ within 
the statistical error.

\begin{figure}[tbp]
\begin{center}
\includegraphics[width=.55\textwidth]{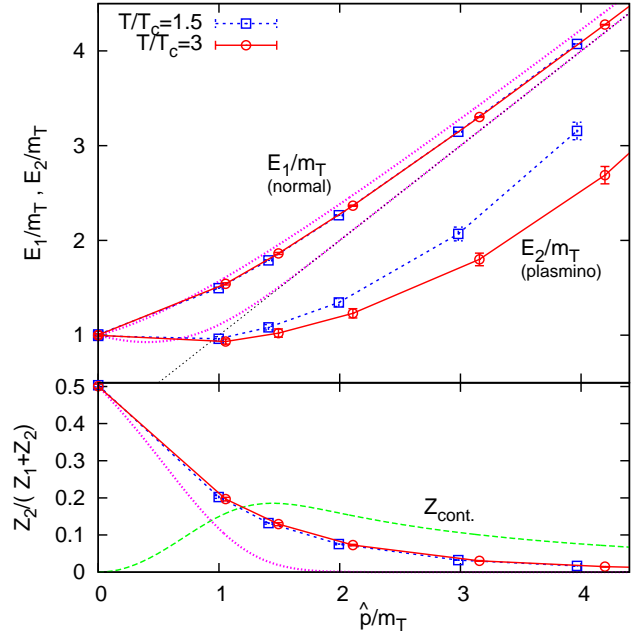}
\caption{
Dependences of the fitting parameters 
$E_1$ and $E_2$ and the ratio $Z_2/(Z_1+Z_2)$ on
the lattice momentum $\hat{p}=(1/a) \sin(pa)$
for $T/T_c=1.5$ and $3$.
See, the text for the explanation of other lines.
}
\label{fig:disp}
\end{center}
\end{figure}

Next, we set $m_0=0$ and analyze the momentum dependence 
of the excitation spectra at nonzero momentum using the 
decomposition given in Eq.~(\ref{eq:rho^P}).
In Fig.~\ref{fig:disp} we show the momentum dependence 
of $E_1$ and $E_2$ normalized by $m_T$ and 
$Z_2/(Z_1+Z_2)$ for $T/T_c=1.5$ and $3$.
The horizontal axis represents the momentum of free Wilson 
fermions on the lattice, $\hat{p}=(1/a) \sin pa$, normalized 
by $m_T$.
The figure shows that $E_1>E_2$ is satisfied in accordance
with the relation between the normal and plasmino dispersions
at asymptotically high temperature.
The figure also shows that the value of $E_2$ 
for the lowest non-zero momentum, $p_{\rm min} = 2\pi 
(N_\tau/N_\sigma)T$, is significantly lower than $m_T$.
Provided that the value of $E_2$ in our two-pole ansatz 
represents the dispersion relation of the plasmino mode,
this result serves as direct evidence for the existence of 
the plasmino minimum in the non-perturbative analysis.
One, however, has to be careful with this interpretation
because the Euclidean correlator is insensitive to the 
spectral function at low energy, $|\omega|\lesssim T$, and 
analysis of the spectrum in this energy range has large 
uncertainty \cite{KKKS}.

\section{Quark self-energy}

In order to investigate dynamical properties of the system 
in lattice simulations, one has to take the analytic 
continuation from a Euclidean correlation function 
obtained on the lattice to a real-time propagator.
This analytic continuation, however, is a famous ill-posed
problem, because one has to infer the real-time propagator, 
which is a continuous function, from finite and noisy data 
obtained on the Monte Carlo simulations.
In previous sections, we have used an ansatz 
Eq.~(\ref{eq:Stau-rho}) for the spectral function
to avoid this difficulty.
Although such an analysis would be convenient to understand 
a qualitative structure of the spectrum, details 
of the spectrum is not accessible.
Even if one uses the maximum entropy analysis which infers
the spectral function without introducing an ansatz, the 
resulting spectral image has uncertainty in the analyses of 
lattice correlators with typical statistics.
To make the analytic continuation more robust, 
therefore, it is desirable to have a different formula 
which relates real- and imaginary-time functions besides
Eq.~(\ref{eq:Stau-rho}).

Here, we propose to exploit the {\it inverse} propagator
for this purpose. 
The inverse of the retarded quark propagator, $S^R(\omega,p)$, 
is written as
\begin{align}
\left[S^R(\omega,p)\right]^{-1} 
= \left[S^R_0(\omega,p)\right]^{-1} - \Sigma^R(\omega,p) ,
\label{eq:SD}
\end{align}
where $S^R_0(\omega,p)$ and $\Sigma^R(\omega,p)$ denote 
the retarded free-quark propagator and self-energy, respectively.
Let us first derive a formula like Eq.~(\ref{eq:Stau-rho})
relating $[S^R(\omega,p)]^{-1}$ to a Euclidean function.
For this purpose we first remark that $[S^R(\omega,p)]^{-1}$ 
is analytic in the upper-half complex-energy plane, $\mathbb{C}^+$,
as well as $S^R(\omega,p)$. This statement is easily verified by 
the fact that $\Sigma^R(\omega,p)$ is analytic in $\mathbb{C}^+$ 
by definition.
Using this property of $[S^R(\omega,p)]^{-1}$ and Kramers-Kronig 
relation, by taking a similar procedure to derive 
Eq.~(\ref{eq:Stau-rho}) one arrives at a formula which connects 
the inverse retarded propagator to a Euclidean function
\begin{align}
\left[\tilde S\right]^{-1}(\tau,p) 
= \int_{-\infty}^\infty d\omega
\frac{ \e^{ (1/2 -\tau T) \omega/T }}{ \e^{\omega/2T} + \e^{-\omega/2T} }
\mbox{Im} \left[S^R(\omega,p)\right]^{-1} 
= -\int_{-\infty}^\infty d\omega
\frac{ \e^{ (1/2 -\tau T) \omega/T }}{ \e^{\omega/2T} + \e^{-\omega/2T} }
\mbox{Im} \Sigma^R(\omega,p),
\label{eq:S-Sig}
\end{align}
where $[\tilde S]^{-1}(\tau,p)$ is the inverse of 
the Matsubara propagator $\tilde S(\tau,p)$.
Remark, however, that Eq.~(\ref{eq:S-Sig}) cannot apply to 
$\tau=0$.
In the last equality of Eq.~(\ref{eq:S-Sig}) we have used the 
fact that $[S_0(\omega,p)]^{-1}$ is real\footnote{
On the lattice, however, $[S_0(\omega,p)]^{-1}$ takes
the imaginary part as a discretization effect.},
and hence $\mbox{Im} [S(\omega,p)]^{-1} 
= - \mbox{Im} \Sigma^R(\omega,p)$.

The inverse propagator $[\tilde S]^{-1}(\tau,p)$ is calculated by 
inverting the correlation function $\tilde S(\tau,p)$ obtained 
on the lattice.
Since $\tilde S(\tau,p)$ is block-diagonal in frequency space,
this inversion is most conveniently taken 
after the Fourier transformation, {\it i.e.}
\begin{align}
\left[\tilde S\right]^{-1}(\tau,p)
= T \sum_n \left[\tilde S(i\omega_n,p)\right]^{-1} e^{ -i\tau\omega_n},
\end{align}
with Matsubara frequency $\omega_n = (2n+1)\pi T$, where
$[~~]^{-1}$ in the r.h.s. represents an inverse of $4\times4$ matrix
for Dirac indices, while that in the l.h.s.
means the inverse of the whole propagator.
Since Eq.~(\ref{eq:S-Sig}) has the same form as 
Eq.~(\ref{eq:Stau-rho}), once $[S]^{-1}(\tau,p)$ is constructed
one can infer the real-time self-enegy $\mbox{Im} \Sigma(\omega,p)$ 
with the same techniques to analyze the spectral function with 
Eq.~(\ref{eq:Stau-rho}), such as the maximum entropy method.

Remarks in constructing $[\tilde S]^{-1}(\tau,p)$ are in order here.
First, $[\tilde S]^{-1}(\tau,p)$ is not the thermal average of 
the fermion matrix $K = i D\hspace{-2.3mm}/ - m$. This is because 
the propagator in the l.h.s. of Eq.~(\ref{eq:SD}) is 
defined by the thermal average of $K^{-1}$. 
The thermal average thus must be taken for the inverse of $K$.
Second, to obtain $[\tilde S]^{-1}(\tau,p)$ one needs all elements 
of $\tilde S(\tau,p)$ including the value at $\tau=0$. 
For correlators having a positive mass dimension, $\tilde S(\tau,p)$ 
is singular at $\tau=0$ in the continuum limit, and 
one must exclude this point from the analysis.
The quark correlator, on the other hand, has zero mass 
dimension and takes a finite value even at the origin. 
No difficulty thus in principle arises with the use of this point.
The use of $\tilde S(\tau,p)$ near the source, however, is
troublesome because correlators with small $\tau/a$ receive strong 
lattice artifacts due to the overlap of operators on the lattice.
We will later investigate this effect by directly analysing 
$[\tilde S]^{-1}(\tau,p)$.

The analysis of $[\tilde S]^{-1}(\tau,p)$ and 
${\rm Im}\Sigma^R(\omega,p)$ has more advantages.
First, in the perturbative analysis of quark propagator
one must first calculate the self-energy.
In this sense, the self-energy is more useful and fundamental 
quantity than the propagator in the analytic studies.
If one wants to compare the analytic result with the lattice one,
the comparison is usually made in terms of the propagator.
The data of the inverse propagator, $[\tilde S]^{-1}(\tau,p)$,
on the other hand, enables to make this comparison 
in terms of the self-energy.
Next, microscopic physics behind the spectral properties would 
become more apparent through the analysis of 
$\mbox{Im}\Sigma(\omega,p)$.
Because the value of $\mbox{Im}\Sigma(\omega,p)$ is related to
elementary processes via the optical theorem, one can give a 
direct interpretations to the structure of $\mbox{Im}\Sigma(\omega,p)$. 
Of course, $\rho(\omega,p)$ and $\mbox{Im}\Sigma(\omega,p)$ 
should have one-to-one correspondence under a given regularization,
and in principle one of these functions must contain all information
of excitation properties.
In lattice simulations, however, one cannot determine the 
real-time function definitely. 
The analysis of a different function thus can make different
physics more apparent.

Because the same correlator as in Eq.~(\ref{eq:Stau-rho}) 
is used in the construction of $[\tilde S]^{-1}(\tau,p)$,
one may think that Eq.~(\ref{eq:S-Sig}) does not provide any
new information on the real-time function besides 
Eq.~(\ref{eq:Stau-rho}).
We, however, remark that Eq.~(\ref{eq:S-Sig}) encodes 
the analyticy of the {\it inverse} propagator in $\mathbb{C}^+$,
which is not taken into account in Eq.~(\ref{eq:Stau-rho}).
An appropriate use of Eq.~(\ref{eq:S-Sig}) thus should enable us
to extract additional information on the real-time function.

\begin{figure}[tbp]
\begin{center}
\includegraphics[width=.6\textwidth]{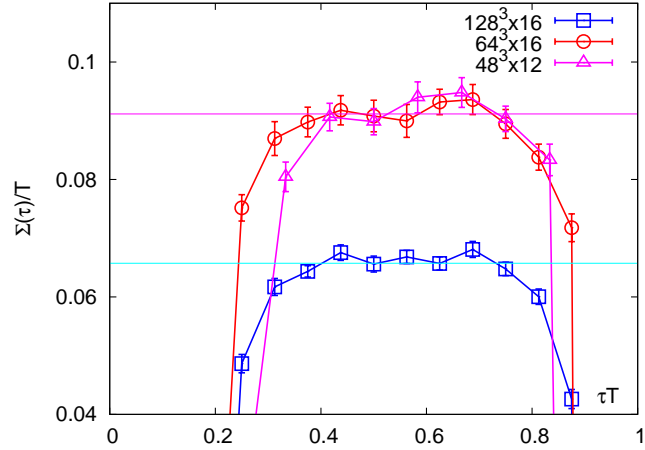}
\caption{
The quark self-energy in Euclidean space as a function of
imaginary time $\tau$ for $T/T_c=3$.
}
\label{fig:selfe}
\end{center}
\end{figure}

In Fig.~\ref{fig:selfe}, we show the inverse quark propagator
$[\tilde S]^{-1}_0(\tau,0)={\rm Tr}[[\tilde S]^{-1}(\tau,0)\gamma^0]$ 
in imaginary time in the chiral limit for $T=3T_c$, with 
several values of $N_\sigma$ and $N_\tau$, which are 
constructed from the same quark correlators 
analyzed in the previous sections \cite{KK,KKKS}.
Errorbars are estimated by the jackknife analysis.
In the figure, we also show the values of 
$[\tilde S]^{-1}_0(\tau,0)$ corresponding to the spectral function 
estimated by the two-pole ansatz Eq.~(\ref{eq:2pole}) by
the thin-solid lines: The spectrum in the chiral limit,
$\rho(\omega,0)= Z( \delta(\omega-m_T)+\delta(\omega+m_T) )$,
corresponds to ${\rm Im}[S(\omega,0)]^{-1}_0 
= {\rm Im} {\rm Tr}[\Sigma(\omega,0)\gamma^0] 
= m_T^2/(TZ) \delta(\omega)$.
One finds from the figure that $[S]^{-1}_0(\tau,0)$ is consistent 
with the prediction of the two-pole ansatz within statistics
near $\tau T=0.5$.
On the other hand, $[\tilde S]^{-1}_0(\tau,0)$ shows a strong 
deviation from the constant near the source.
Similar results are obtained for nonzero $m_0$ and $p$.

The fact that the values of $[\tilde S]^{-1}_0(\tau,p)$ coincide 
with the ones predicted by the pole ansatz around $\tau T=0.5$
would support (1) the validity of the pole ansatz for the quark 
spectrum, and (2) the relevance of the evaluation of 
${\rm Im}\Sigma(\omega,p)$ with Eq.~(\ref{eq:S-Sig}).
The large deviation from the constant near the source would be 
attributed to the distortion effects; in fact, the deviation 
is more prominent on the course lattice.
The errorbars and the distortion effect with the present lattice 
data are too large to constrain the form of the spectral function 
with Eq.~(\ref{eq:S-Sig}).
Much finer lattice and higher statistics are needed 
to proceed the analysis of the self-energy with 
Eq.~(\ref{eq:S-Sig}) further.

To summarize, in this report we analyzed the spectral 
properties of quarks on the quenched lattice with size
$N_\sigma^3\times N_\tau=128^3\times16$.
We found that the spatial volume dependence of the quark thermal 
mass defined by the pole ansatz tends to converge on this volume.
The dispersion relations of the normal and plasmino modes are
investigated within the two-pole ansatz. 
The result indicates the existence of a minimum of the plasmino
dispersion at nonzero momentum.
We also introduced attempts to analyze the quark self-energy.

This proceedings is based on the collabolation done with 
O.~Kaczmarek, F.~Karsch, W.~Soeldner, M.~Asakawa, and S.~Takotani.
Numerical simulations of this study have been performed on the 
BlueGene/L at the New York Center for Computational Sciences 
(NYCCS).


\begin{thebibliography}{99}

\bibitem{LeBellac}
  M.~Le~Bellac, {\it Thermal Field Theory}
  (Cambridge University Press, Cambridge, England 1996).

\bibitem{Fries:2003kq}
  R.~J.~Fries, B.~Muller, C.~Nonaka and S.~A.~Bass,
  Phys.\ Rev.\  C {\bf 68}, 044902 (2003)
  [arXiv:nucl-th/0306027].

\bibitem{KK}
  F.~Karsch and M.~Kitazawa,
  Phys.\ Lett.\  B {\bf 658}, 45 (2007)
  [arXiv:0708.0299 [hep-lat]];
  Phys.\ Rev.\  D {\bf 80}, 056001 (2009)
  [arXiv:0906.3941 [hep-lat]].

\bibitem{KKKS}
  O.~Kaczmarek, F.~Karsch, M.~Kitazawa, and W.~Soeldner,
  in preparation.

\bibitem{BBS92}
  G.~Baym, J.~P.~Blaizot and B.~Svetitsky,
  Phys.\ Rev.\ D {\bf 46}, 4043 (1992).

\bibitem{KKN}
  M.~Kitazawa, T.~Kunihiro and Y.~Nemoto,
  Phys.\ Lett.\ B {\bf 633}, 269 (2006)
  [arXiv:hep-ph/0510167];
  Prog.\ Theor.\ Phys.\  {\bf 117}, 103 (2007)
  [arXiv:hep-ph/0609164];
  M.~Kitazawa, T.~Kunihiro, K.~Mitsutani and Y.~Nemoto,
  Phys.\ Rev.\  D {\bf 77}, 045034 (2008)
  [arXiv:0710.5809 [hep-ph]].

\end{thebibliography}
\end{document}